%% file: main.tex
\documentclass[10pt]{article} 
\usepackage[preprint]{tmlr}

\input{math_commands.tex}

\usepackage{hyperref}
\usepackage{url}

\usepackage{booktabs}
\usepackage{color}
\usepackage{url}
\usepackage{amsfonts}
\usepackage{graphicx}
\usepackage{subcaption}
\usepackage{amsmath}

\title{Conversational Speech Separation: an Evaluation Study for Streaming Applications}

\author{\name Giovanni Morrone \email g.morrone@univpm.it \\
    \name Samuele Cornell \email s.cornell@pm.univpm.it \\     
    \addr Department of Information Engineering \\
    Università Politecnica delle Marche, Ancona, Italy
    \AND
    \name Enrico Zovato \email enrico.zovato@pervoice.it \\
    \addr PerVoice S.p.A., Trento, Italy \\
    \AND
    \name Alessio Brutti \email brutti@fbk.eu \\
    \addr Fondazione Bruno Kessler, Trento, Italy \\
    \AND
    \name Stefano Squartini \email s.squartini@univpm.it \\
    \addr Department of Information Engineering \\
    Università Politecnica delle Marche, Ancona, Italy
    }

\def\month{MM}  
\def\year{YYYY} 
\def\openreview{\url{https://openreview.net/forum?id=XXXX}} 

\begin{document}

\maketitle

This paper was first published as Preprint 10562 at the 152 AES Spring Convention in the Hague May 2022. Available at: \url{https://www.aes.org/e-lib/browse.cfm?elib=21675}.

\begin{abstract}
Continuous speech separation (CSS) is a recently proposed framework which aims at separating each speaker from an input mixture signal in a streaming fashion. 
Hereafter we perform an evaluation study on practical design considerations for a CSS system, addressing important aspects which have been neglected in recent works. In particular, we focus on the trade-off between separation performance, computational requirements and output latency showing how an offline separation algorithm can be used to perform CSS with a desired latency.
We carry out an extensive analysis on the choice of CSS processing window size and hop size on sparsely overlapped data. We find out that the best trade-off between computational burden and performance is obtained for a window of $5$ s.
\end{abstract}

\section{Introduction}
\label{sec:intro}

The goal of speech separation is to extract every speech source in an input mixture signal. Speech separation is a fundamental task in many real-world applications such as hearing aids, robust automatic speech recognition and mobile communication to name a few \cite{vincent2018audio}. Indeed, as a natural phenomenon in human interactions, overlapping speech is present in varying degrees in most conversations.

In the last few years, speech separation has dramatically improved using deep learning techniques \cite{wang2018supervised}, and deep learning driven systems are now the mainstream approach for speech and, in general, source separation.

The consistent progress of speech separation systems has been demonstrated by steady improvements of the \textit{scale-invariant signal-to-distortion ratio} metric (SI-SDR) \cite{leroux2019sdr} on the WSJ0-2mix dataset \cite{hershey2016deep}, which has been the de-facto benchmark dataset for speech separation. However, this dataset only considers the fully-overlapped speech case, which does not match real-world scenarios. In fact, the overlap ratio between multiple speakers is usually between $10$\% and $20$\% for multiparty conversations \cite{cetin2006analysis, cornell2022overlapped}.

For this reason, there is a growing interest in \textit{continuous speech separation} (CSS), which aims at generating multiple audio streams, one for each speaker, from an input audio stream where the overall speaker overlap is low and the speaker activations are sparse (as happens e.g. in a meeting scenario with multiple speakers) \cite{chen2021continuous, han2021continuous, li2021dual}. Generally, the majority of single-channel CSS systems focus on offline processing. The relationship between latency, resource usage and separation performance has not been sufficiently analyzed yet. However, in streaming applications, such as live captioning in e.g. teleconferencing, low latency and power consumption are fundamental concerns.

In this work, we try to bridge the gap between recently proposed separation algorithms and their application on real-world settings. To carry out our analysis, we consider a dataset of real phone conversations and use it to generate synthetic datasets both for training and testing purposes. We employ a single-channel offline speech separator, the dual-path recurrent neural network (DPRNN) \cite{luo2020dual}, trained on fixed-length segments which was demonstrated to reach promising performance for the CSS task \cite{li2021dual}. Firstly, we study how different speaker overlap ratios affect the separation performance using different processing windows. Then, we show how an offline separator can be used to perform CSS in a causal setting with a fixed latency and processing window, this latter defining the computational cost. We provide an extensive evaluation of the performance with different latencies and processing windows, that can be useful in the choice of the right trade-off between separation performance and computational cost.

The paper is organized as follows. In the following Subsection we present a summary of the related work about speech separation. In Section \ref{sec:sys_descr} we briefly introduce the general CSS framework and its conversion from offline to online/causal processing. The experimental setup is presented in Section \ref{sec:exp_setup}. Results are shown and discussed in Section \ref{sec:res}. Finally, in Section \ref{sec:conclusions} we draw conclusions and outline possible future work.

\subsection{Related Work}
\label{ssec:intro/rel_work}

In this subsection we review speech separation algorithms based on deep learning, which are the most related to our work.

Hershey et al. \cite{hershey2016deep} proposed \textit{deep clustering} (DC), which used a recurrent network with an objective function that was invariant to the number of speakers and their permutation. In \cite{kolbaek2017multitalker} the \textit{permutation invariant training} (PIT) approach was proposed, which solved the speaker permutation problem by searching the best combination between output and target channels during model training. The DC and the PIT strategies enabled a major breakthrough in the field and several subsequent works introduced extensions that improved various aspects. Chen et al. \cite{chen2017deep} presented a method, named \textit{deep attractor network} (DANet), similar to DC which can be trained in an end-to-end fashion. In another work \cite{wang2018alternative}, the authors compared a variety of alternative objective functions and combined DC with mask-inference networks in a multi-task learning (MTL) framework. All the mentioned works attempted to generate a mask for each speaker in the Short-Time Fourier Transform (STFT) magnitude time-frequency (TF) representation. The current state-of-the-art employed an encoder-masker-decoder framework \cite{luo2020dual, luo2018tasnet, luo2019conv, dptn, subakan2020attention}. In these works, separation models directly operated in the mixture waveform domain by learning analysis and synthesis filterbanks in place of the STFT, and using a deep neural network (DNN) to estimate a per-speaker mask in this learned TF representation to separate the speakers signals.

Yoshioka et al. \cite{yoshioka2018recognizing} were the first to propose a multi-channel CSS approach and apply it on real-world meetings. They combined a bidirectional long short-term memory (BLSTM) network \cite{graves2013hybrid} with a improved mask-based beamformer. A follow-up work \cite{chen2020continuous} introduced a new dataset, named LibriCSS, and an evaluation protocol to test CSS algorithms. It showed that even when the separation was aided by a seven-channel minimum variance distortionless response (MVDR) beamformer \cite{erdogan2016improved} the speech recognition degradation on overlapped speech segments was not trivial. Moreover, they observed that the word error rate (WER) sometimes increased when only one speaker was active. These works only focused on multi-channel approaches, ignoring the single-channel scenario that is much more challenging and desirable from an application standpoint. In \cite{chen2021continuous} a system based on the Conformer \cite{gulati2020conformer} model was proposed for CSS, and was shown to outperform previous multi-channel systems. Han et al. \cite{han2021continuous} leveraged the sparsity of overlapped regions to derive robust speaker embeddings of individual talkers from single-speaker segments. The speaker embeddings were included in a speaker inventory used in an adaptation layer before the separation module. The use of dual-path modeling for speech separation in meetings was studied by Li et al. \cite{li2021dual}. In this work the authors replaced the BLSTMs in DPRNN blocks with transformer layers \cite{vaswani2017attention}. The updated model outperformed the baseline reported in \cite{chen2020continuous} and reduced the amount of computations by 30\%.
These works only focused on offline processing. A notable exception is \cite{li2021dual} where a comparison between the performance of the offline DPRNN and an online version of the same model is presented. Results showed that only for short local processing window sizes (i.e., $0.8$ and $1.6$ s) the offline model outperformed the online version for several low-overlap conditions.

\begin{figure*}[t]
\begin{center}
\includegraphics[width=1.\textwidth]{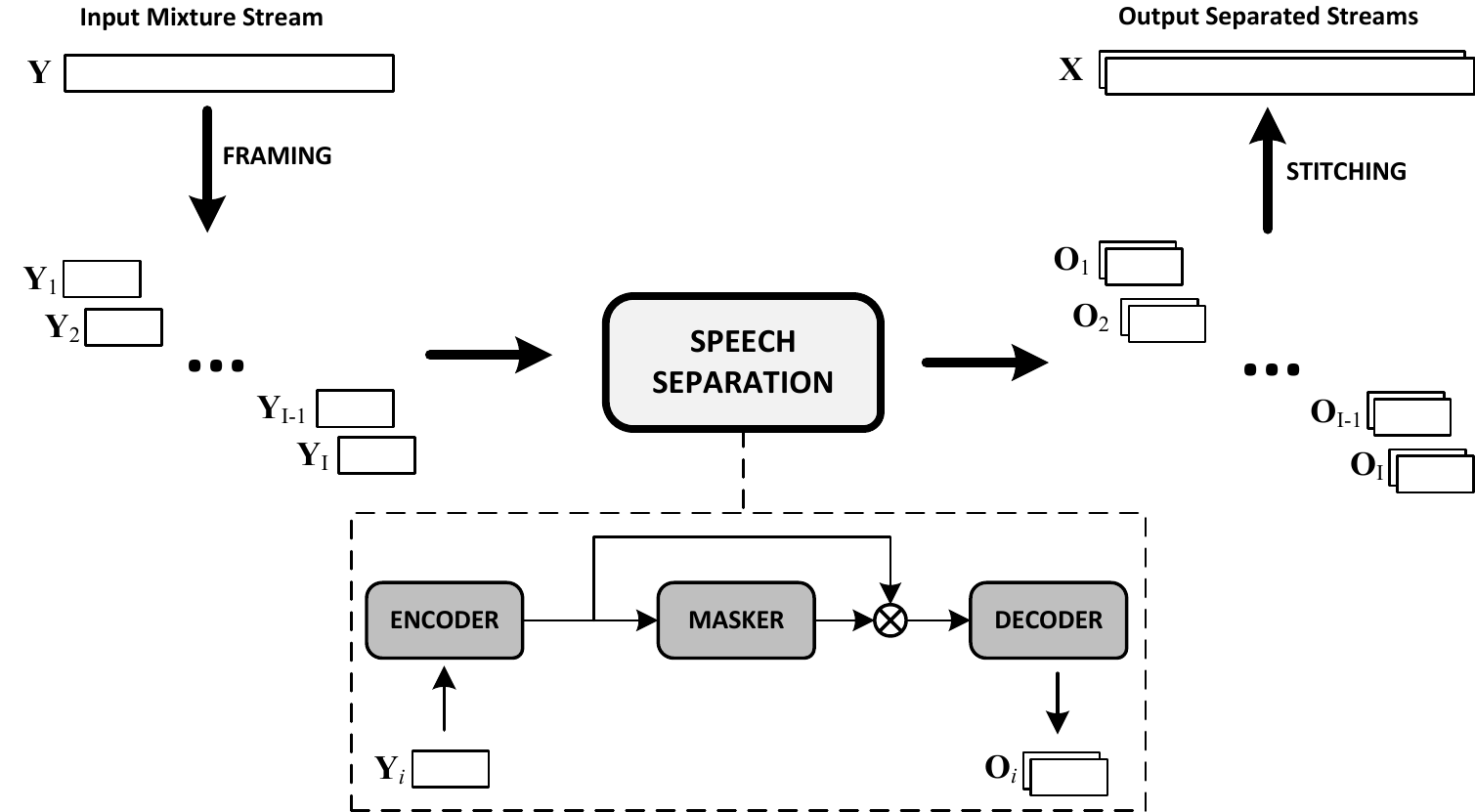}
\caption{General diagram of the CSS architecture.}
\label{fig:sys_descr/architecture}
\end{center}
\end{figure*}

\section{System Description}
\label{sec:sys_descr}

\subsection{CSS Framework}
\label{ssec:sys_descr/css}

The CSS framework \cite{li2021dual} consists of three modules: framing, separation, and stitching. A diagram of the architecture of the system is depicted in Fig. \ref{fig:sys_descr/architecture}.

The input of the system is the mixture audio stream. We denote the input stream with $\mathbf{Y} \in \mathbb{R}^{1 \times T}$, where $T$ is the number of audio samples in the waveform. 

\subsubsection{Framing Module}
\label{sssec:sys_descr/css/framing}

In the framing stage we apply a windowing operation which splits $\mathbf{Y}$ into $I$ overlapped frames $\mathbf{Y}_i \in \mathbb{R}^{1 \times W}, i = 1, \dots, I$, with $I = \lceil \frac{T}{H} \rceil$, where $W$ and $H$ are the window and hop size, respectively.
Then, separation is applied on each frame $\mathbf{Y}_i$, and generates separated output frames $\mathbf{O}_i \in \mathbb{R}^{C \times W}$, where $C$ is the number of output channels. In this work, $C$ is fixed to $2$, meaning that we assume that the maximum number of speakers in input mixtures is $2$. Two-speakers overlap is also the most common scenario in real-world meetings and conversations \cite{cetin2006analysis, cornell2022overlapped}.  

\subsubsection{Speech Separation Module}
\label{sssec:sys_descr/css/ss}

Although we can use every speech separation algorithms trained with PIT as the separation module, in this work we choose to use DPRNN \cite{luo2020dual} to carry out our analysis as it is a well established state-of-the-art separation model with a good trade-off between computational requirements and performance.

The DPRNN consists of three parts: an encoder, a masker and a decoder - see the dotted box in Fig. \ref{fig:sys_descr/architecture}. The encoder converts the input waveform in an 2-D STFT-like adaptive representation using an 1-D convolutional layer. The masker is fed with the 2-D input and estimates $C$ masks in the same format. The 2-D input is multiplied with the masks to obtain the separated 2-D outputs. Finally, an 1-D transposed convolutional layer decoder converts the masked 2-D representations back to waveform domain. 
The masker uses a clever windowing strategy and several blocks of recurrent layers to model long sequential inputs. The input sequence is split into shorter chunks which are processed by an intra-chunk and an inter-chunk block. The intra-chunk block processes the local chunks independently. Then, the outputs from several chunks are aggregated to perform utterance-level processing. Such approach allows each of the processing block to only receive a small fraction of the input feature sequence and simplify the optimization during training. A combination of intra- and inter-chunk blocks is denoted as a DPRNN block, and several DPRNN blocks can be stacked to obtain deeper networks. Both intra- and inter-chunks blocks are based on BLSTMs \cite{graves2013hybrid}.

\subsubsection{Stitching Module}
\label{sssec:sys_descr/css/stiching}

Since the DPRNN model is trained with a permutation-free objective function \cite{kolbaek2017multitalker}, it is not guaranteed that a speaker is always placed in the same channel in every frame. The stitching module aligns channels of two separation outputs $\mathbf{O}_i$ and $\mathbf{O}_{i+1}$ according to a similarity metric computed on the overlapped part of consecutive frames. The size of the overlapped part is $W - H$. We use the cross-correlation as similarity metric to reorder the frames. The final output stream $\mathbf{X} \in \mathbb{R}^{C \times T}$ is generated by an overlap-add operation with Hanning window. The use of Hanning window instead of rectangular (as used in e.g. \cite{yoshioka2018recognizing}) helps smooth the signal at edges and produces better estimations.

\subsection{From Offline to Online/Causal Processing}
\label{ssec:sys_descr/online_proc}

In the CSS framework, latency $t_{l}$ is equal to the sum of the length of processing window $t_W = \frac{W}{sr}$, where $sr$ is the audio sampling rate, and processing time $t_{proc}$ for a single frame:
\begin{equation}
    t_{l} = t_W + t_{proc}.
\end{equation}
$t_{proc}$ is not constant as it depends on both model complexity and available computational power. On the other hand, $t_W$ is only defined by the choice of the window size $W$, then represents the minimum latency we can achieve for a given $W$.

\begin{figure}[t]
\begin{center}
\includegraphics[width=.8\columnwidth]{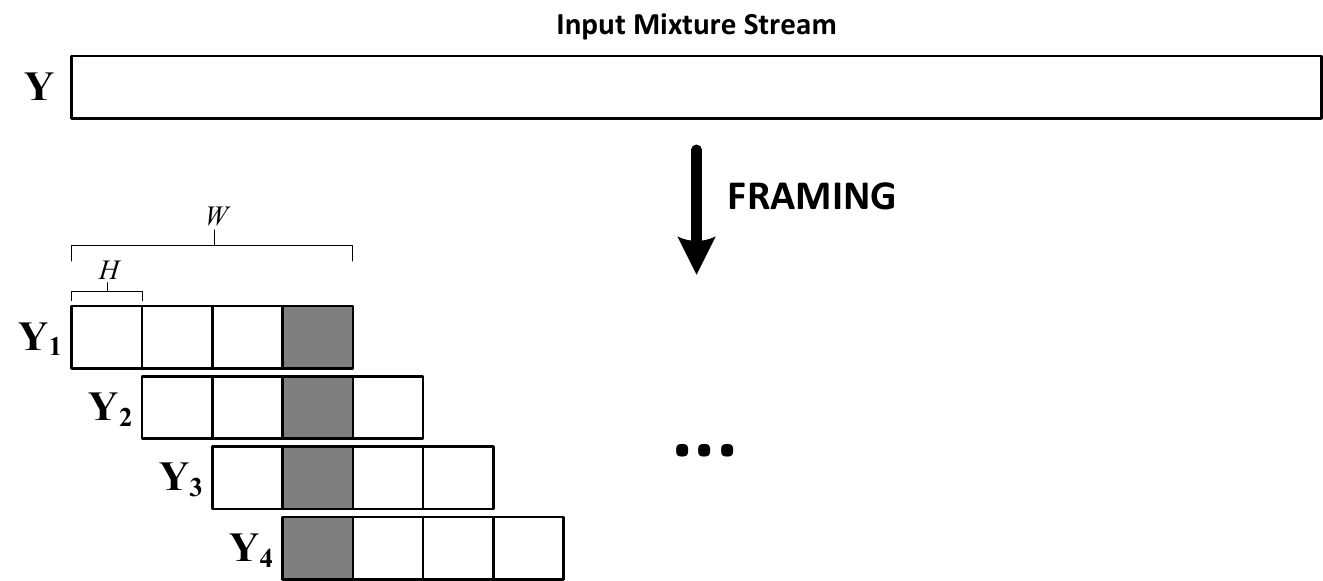}
\caption{Example of windowing with $75$\% overlap between consecutive frames. The gray boxes represent the same segment included in consecutive frames in different positions.}
\label{fig:sys_descr/win_seg}
\end{center}
\end{figure}

Fig. \ref{fig:sys_descr/win_seg} shows the framing stage when the overlapping degree between frames is set to $75$\%. In this case, the hop size is $4$ times smaller that the window size. Then, a given segment of length $H$ is included in $4$ consecutive frames, which generates $4$ different outputs for such segment. 

However, all outputs are not necessarily required. We can relax this constraint and only exploit some of outputs to emit the final estimation. The conversion from a generic offline separator to an online one with desired latency is done by simply reducing the hop size and the number of segments $n_{seg}$, from 1 to $\frac{W}{H}$, exploited to generate the final output streams.
In this way, we can reduce the minimum latency from $t_W$ to $\frac{n_{seg} * H}{sr}$. When $n_{seg} = 1$, only the segment in the first frame is used which results in the minimum latency $t_H = \frac{H}{sr}$ we can achieve for a given hop size $H$.

\section{Experimental Setup}
\label{sec:exp_setup}

\subsection{Dataset}
\label{ssec:exp_setup/dataset}

The objective of this work is to carry out an analysis of speech separation on natural and spontaneous conversations. We use the Fisher Corpus Part $1$ \cite{cieri2004fisher} as both training and test data. This dataset consists of $5850$ phone conversations between two participants, sampled at $8$ kHz of both native and non-native English speakers. The total amount of active speech is about $1012$ h. Transcriptions and time boundaries of each speaker utterances are provided. Audio from the two speakers are recorded on two separated channels, and human and environmental noises can be present. Such noises are annotated in the transcriptions. Since, in this work, we focus on speech separation only and not in combined speech separation and enhancement we discard noisy utterances. That reduces the amount of available speech data to $776$ h. The overlap ratio between the two speakers amounts to $14.6$\% and $8.3$\% for all data and data without noises, respectively. This difference is mainly due to the presence of human non-speech sounds from one speaker (i.e., laugh) when the other one is talking.

For our purposes, we create a mixed-speech version of the Fisher Corpus. Utterances from two channels belonging to different conversations are mixed together with volume levels sampled from uniform distribution between $-33$ dB to $-25$ dB as in \cite{cosentino2020librimix} to introduce additional randomness in speech data. We discard utterances shorter than $3$ seconds and the minimum overlap ratio is set to $80$\%, as the state-of-the-art speech separation models demonstrated to work properly when trained with fully overlapped speech. To ensure that the separation system is speaker-independent, and performs reliably independently of speaker identity/gender, the conversations are divided in $3$ disjoint sets which are associated to training, validation, and test. In this way, a given speaker can only appear in one of three sets. The synthetic dataset is generated by drawing $30000$ ($44.8$ h), $3000$ ($4.7$ h) and $3000$ ($5$ h) utterances for training, validation and test sets, respectively.

In addition, we create sparsely overlapped datasets, that mimic real-world conversational scenarios similar to SparseLibriMix \cite{cosentino2020librimix}. A training set of $6000$ ($15.9$ h) conversations with $0$\%, $10$\% and $20$\% overlap is used jointly with the fully overlapped training set. Moreover, to test the performance by varying the overlap degree, we generate several sparse test sets. Each conversation consists of alternating utterances from the two speakers. We pick utterances from $2$ to $5$ s of each speaker until we reach a minimum length of $15$ s. A first dataset is generated by concatenating all the utterances without overlapping them. Each utterance is separated by a $50$ ms-long pause. From this overlap-free dataset we create different sparsely overlapped versions using the same utterances and overlapping them by $10$\%, $20$\%, $40$\%, $60$\%, $80$\%, and $100$\% of the time. For each overlap ratio $500$ conversations are created, resulting in $3500$ ($12.3$ h) examples.

\subsection{Model Configuration}
\label{ssec:exp_setup/model_conf}

The kernel size, the stride, and the number of filters of the encoder and the decoder are set to $16$, $8$, and $64$, respectively. For the masker we use $6$ DPRNN blocks. BLSTMs \cite{graves2013hybrid} with $128$ hidden units are used for both intra- and inter-chunk blocks. The chunk size is set to $100$ with $50$\% overlap. A sigmoid activation layer is used to limit masks values in the $(0,1)$ interval.

\subsection{Training and Optimization}
\label{ssec:exp_setup/opt}

The experiments are conducted using the Asteroid \cite{pariente20} toolkit, which is based on the PyTorch \cite{paszke2019pytorch} deep learning framework. We train the main speech separation model for a maximum of $200$ epochs on $3$-seconds long segments. The fixed-length segments are extracted randomly from examples of the fully overlapped training dataset. The model is trained with utterance-level PIT \cite{kolbaek2017multitalker} to maximize the SI-SDR metric \cite{leroux2019sdr}. Early stopping is applied when the model does not improve on the validation set for $10$ consecutive epochs. To avoid steep gradient fluctuations, gradient clipping with maximum $L_2$-norm of $5$ is applied. We used Adam \cite{kingma2015adam} optimizer setting the initial learning rate to $0.001$. The learning rate is halved when the loss does not decrease for $5$ consecutive epochs.

To improve generalization over sparsely overlapped speech, we apply a fine-tuning using the sparse training set. The set of hyperparameters is the same as above, except for the initial learning rate that is set to $0.0001$.

\section{Results}
\label{sec:res}

Separation performance is measured using the SI-SDR \cite{leroux2019sdr}, which is the most widely used metric to assess the speech separation capability. SI-SDR is defined as:

\vspace{-8mm}

\begin{equation}
\begin{split}
\text{SI-SDR} & = 10 \log_{10} \frac{\left\lVert \mathbf{x}_{scaled} \right\rVert^2}{\left\lVert \mathbf{x}_{err} \right\rVert^2}, \\
\mathbf{x}_{scaled} & = \frac{\langle \mathbf{x}, \hat{\mathbf{x}} \rangle}{\left\lVert \mathbf{x} \right\rVert^2} \mathbf{x}, \\
\mathbf{x}_{err} & = \mathbf{x}_{scaled} - \hat{\mathbf{x}},
\end{split}
\end{equation}

\vspace{-4mm}

where $\mathbf{x}$ and $\hat{\mathbf{x}}$ are the target and estimated single-speaker sources, respectively. $\langle \cdot , \cdot \rangle$ denotes the scalar product, while $\left\lVert \cdot \right\rVert$ is the $L_2$-norm.

\begin{table*}[t]
  \centering
  \resizebox{\textwidth}{!}{
  \begin{tabular}{lc @{\hspace{0.5\tabcolsep}} c @{\hspace{0.5\tabcolsep}} c  @{\hspace{0.5\tabcolsep}} c | c @{\hspace{0.5\tabcolsep}} c @{\hspace{0.5\tabcolsep}} c @{\hspace{0.5\tabcolsep}} c}
    \toprule
    {} &
      \multicolumn{4}{c|}{Model 1 (fully overlap training)} &
      \multicolumn{4}{c}{Model 2 (sparse fine-tuning)} \\ 
     {Overlap ratio} & \multicolumn{1}{c}{1 s} & \multicolumn{1}{c}{3 s} & \multicolumn{1}{c}{5 s} & \multicolumn{1}{c}{15 s} & \multicolumn{1}{c}{1s} & \multicolumn{1}{c}{3 s} & \multicolumn{1}{c}{5 s} & \multicolumn{1}{c}{15 s} \\
      \midrule
    100\% (single-utt.) & 10.35 (6.06) & 13.02 (4.34) & \textbf{13.28} (4.03) & 12.93 (3.98) & 7.80 (6.96) & 13.15 (4.55) & \textbf{13.65} (4.26) & 13.30 (4.31) \\
    \midrule
    0\%   & 1.38 (6.61) & 8.18 (10.27)  & 14.49 (11.04) & \textbf{20.22} (9.39) & -1.43 (7.70) & 25.20 (21.74) & 31.72 (19.50) & \textbf{38.82} (15.63) \\
    10\%  & 3.32 (8.82) & 11.52 (11.04) & 16.44 (10.64) & \textbf{20.56} (9.47) & 3.91 (13.53) & 23.63 (17.48) & 29.01 (15.36) & \textbf{32.76} (12.62) \\
    20\%  & 6.32 (9.69) & 11.90 (10.43) & 16.58 (9.38)  & \textbf{18.90} (8.32)  & 6.11 (12.23) & 18.90 (12.21) & 22.04 (10.75) & \textbf{24.04} (8.99) \\
    40\%  & 7.00 (8.32) & 13.28 (6.66)  & 14.10 (6.22)  & \textbf{15.22} (5.29)  & 6.13 (8.82)  & 14.74 (7.07)  & 16.08 (6.05)  & \textbf{16.69} (5.12) \\
    60\%  & 7.11 (7.83) & 12.77 (6.16)  & 13.71 (5.40)  & \textbf{14.21} (4.78)  & 4.92 (7.99)  & 13.91 (6.31)  & 14.82 (5.55)  & \textbf{15.15} (4.89) \\
    80\%  & 6.65 (7.81) & 12.42 (6.16)  & 13.44 (5.20)  & \textbf{13.80} (4.73)  & 5.06 (7.61)  & 13.60 (5.98)  & 14.39 (5.34)  & \textbf{14.53} (4.84) \\
    100\% & 5.68 (7.40) & 11.93 (5.84)  & 12.70 (4.97)  & \textbf{13.45} (4.15)  & 3.92 (7.13)  & 12.66 (6.19)  & 13.63 (5.11)  & \textbf{13.78} (4.64) \\
    \bottomrule
  \end{tabular}
  }
  \caption{Speech separation results on test sets with different processing windows $W$ ($50$\% overlap) using cross-correlation to reorder consecutive frames at inference. The results are reported in terms of the SI-SDR improvement over the input mixture. We report both the mean and, in parenthesis, standard deviation.}
  \label{tab:res/sep_win}
\end{table*}

\begin{table*}[t]
  \centering
  \resizebox{\textwidth}{!}{
  \begin{tabular}{lc @{\hspace{0.5\tabcolsep}} c @{\hspace{0.5\tabcolsep}} c  @{\hspace{0.5\tabcolsep}} c | c @{\hspace{0.5\tabcolsep}} c @{\hspace{0.5\tabcolsep}} c @{\hspace{0.5\tabcolsep}} c}
    \toprule
    {} &
      \multicolumn{4}{c|}{Model 1 (fully-overlap training)} &
      \multicolumn{4}{c}{Model 2 (sparse fine-tuning)} \\
     {Overlap ratio} & \multicolumn{1}{c}{1 s} & \multicolumn{1}{c}{3 s} & \multicolumn{1}{c}{5 s} & \multicolumn{1}{c}{15 s} & \multicolumn{1}{c}{1s} & \multicolumn{1}{c}{3 s} & \multicolumn{1}{c}{5 s} & \multicolumn{1}{c}{15 s} \\
      \midrule
    100\% (single-utt.) & 12.08 (3.98) & 13.20 (3.77) & \textbf{13.33} (3.90) & 12.94 (3.96) & 9.73 (5.30) & 13.22 (4.32) & \textbf{13.67} (4.13) & 13.31 (4.26) \\
    \midrule 
    0\%   & 21.16 (6.53) & 15.39 (6.61)  & 17.35 (8.04)  & \textbf{20.47} (8.95)  & 36.06 (10.08) & 31.80 (15.37) & 33.99 (16.30) & \textbf{39.26} (14.79) \\
    10\%  & 20.20 (5.89) & 16.52 (7.35) & 18.64 (8.10)   & \textbf{20.67} (9.18)  & 28.81 (8.87)  & 28.46 (12.18) & 30.91 (12.49) & \textbf{32.91} (12.31) \\
    20\%  & 18.23 (5.16) & 16.36 (6.79) & 18.06 (7.39)   & \textbf{19.12} (7.89)  & 21.10 (6.62)  & 22.28 (8.22)  & 23.37 (8.49) & \textbf{24.25} (8.64) \\
    40\%  & 14.49 (3.77) & 14.81 (4.63)  & 15.07 (4.59)  & \textbf{15.37} (4.99)  & 13.48 (5.03)  & 16.32 (4.77)  & 16.72 (4.70)  & \textbf{16.82} (4.81) \\
    60\%  & 13.65 (3.46) & 14.07 (4.18)  & 14.25 (4.28)  & \textbf{14.39} (4.31)  & 11.93 (4.80)  & 15.03 (4.38)  & \textbf{15.50} (4.15)  & 15.24 (4.63) \\
    80\%  & 13.37 (3.40) & 13.85 (4.02)  & \textbf{13.94} (4.18)  & 13.93 (4.35)  & 11.61 (4.71)  & 14.65 (4.23)  & \textbf{15.12} (3.90)  & 14.64 (4.57) \\
    100\% & 12.78 (3.18) & 13.34 (3.75)  & 13.27 (3.98)  & \textbf{13.50} (4.04)  & 10.70 (4.72)  & 13.91 (4.23)  & \textbf{14.36} (3.74)  & 13.88 (4.47) \\
    \bottomrule
  \end{tabular}
  }
  \caption{Speech separation results on test sets with different processing windows $W$ ($50$\% overlap) using oracle information to reorder consecutive frames at inference. The results are reported in terms of the SI-SDR improvement over the input mixture. We report both the mean and, in parenthesis, standard deviation.}
  \label{tab:res/sep_win_oracle}
\end{table*}

\subsection{Offline Evaluation}
\label{ssec:res/offline_eval}
Firstly, we evaluate the CSS system in an offline setting. In this case, we exploit all the available estimations for each segment of length $H$. Thus the latency is equal to the length of the window used in the framing module. The computational cost is the same for all windows, although larger windows requires higher memory usage.

The Table \ref{tab:res/sep_win} reports the results on the test sets for the model trained on fully overlapped speech only (Model 1), and the model fine-tuned on sparsely overlapped speech (Model 2). We test several window lengths, i.e., $1$, $3$, $5$ and $15$ s, with the hop size set to $\frac{W}{2}$ that results in $50$\% overlap between consecutive frames. The Table \ref{tab:res/sep_win_oracle} shows the same results using oracle information from target sources in place of cross-correlation to correctly reorder channels of consecutive output frames. The evaluation on fully overlapped test set with single-utterances is showed in the first row. The performances are similar for the different windows, except for the $1$-second long case, that reaches lower accuracy.

The results on the sparse test set provide more interesting insights. Firstly, larger windows generally perform better, especially for low overlaps. The $1$-second window completely fails for very low overlaps. However, we observe large improvements when using oracle information to reorder consecutive frames. In particular, the SI-SDR improves from $1.28$ to $21.16$ dB and from $3.32$ to $20.20$ dB for the sparse test sets with $0$\% and $10$\% overlap ratios, respectively. This demonstrates that the main source of error is the wrong channel reordering generated by using similarity based on cross-correlation computed on shared portions of consecutive frames. In this case, the size of shared portions, i.e., $0.5$ s, is not large enough to produce accurate similarity estimations. Therefore, it is convenient to use larger windows. The gap between performance with cross-correlation and oracle-based reordering gets smaller when the window size increases. In particular, the gap for the $15$-seconds long window case is basically negligible.

The fine-tuning on sparse speech outperforms the first model for all windows and for all test sets. As expected, we obtain large improvements for low overlap test sets which is highly desirable for real-world scenario where the fraction of overlapped speech is usually below $20$\%.

\begin{figure*}[!ht]
\begin{subfigure}[b]{.5\textwidth}
  \centering
  \includegraphics[width=1.\textwidth]{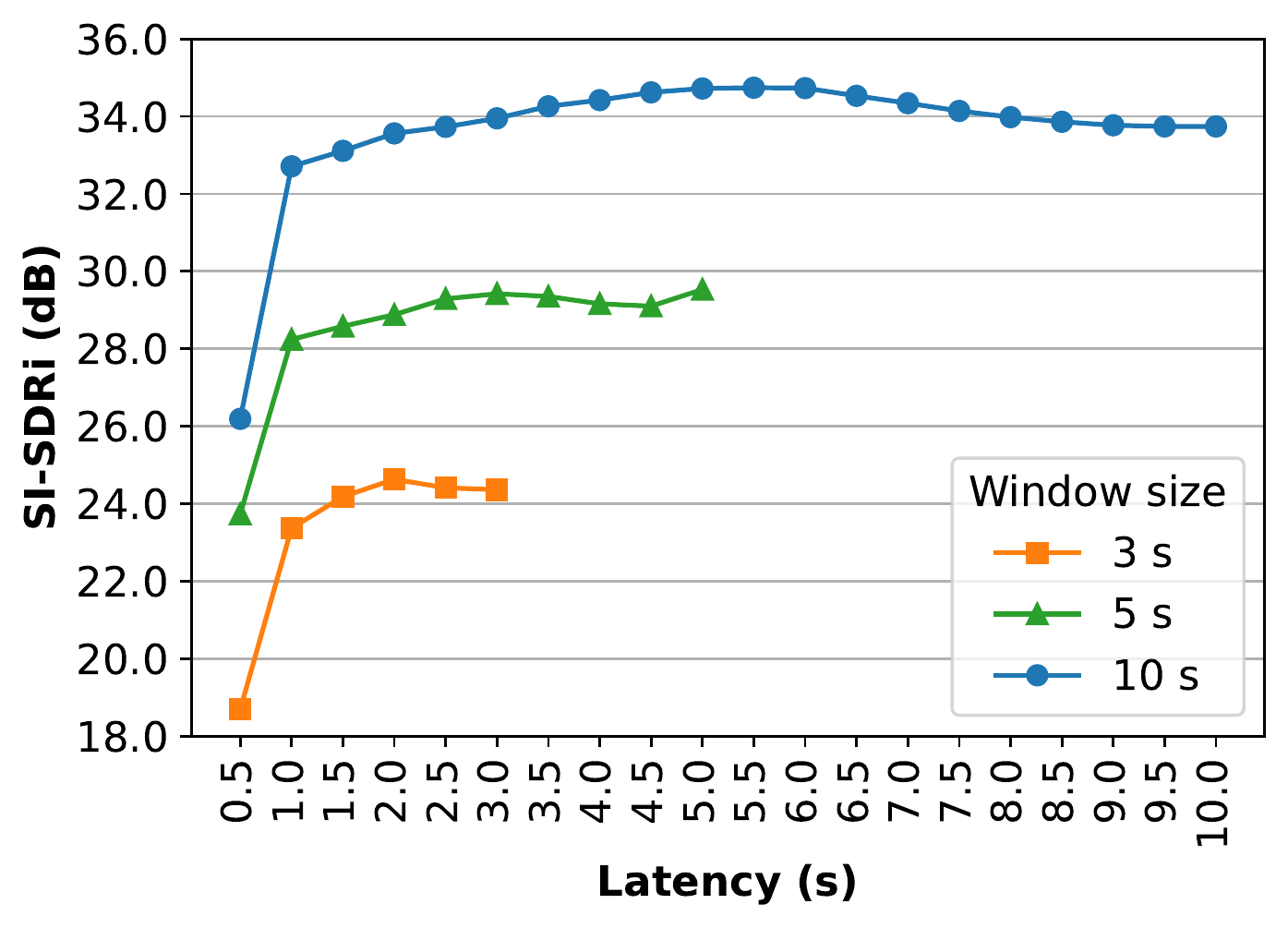}
  \caption{Overlap ratio: $0$\%}
  \label{fig:res/causal_analysis/0}
\end{subfigure}
\begin{subfigure}[b]{.5\textwidth}
  \centering
  \includegraphics[width=1.\textwidth]{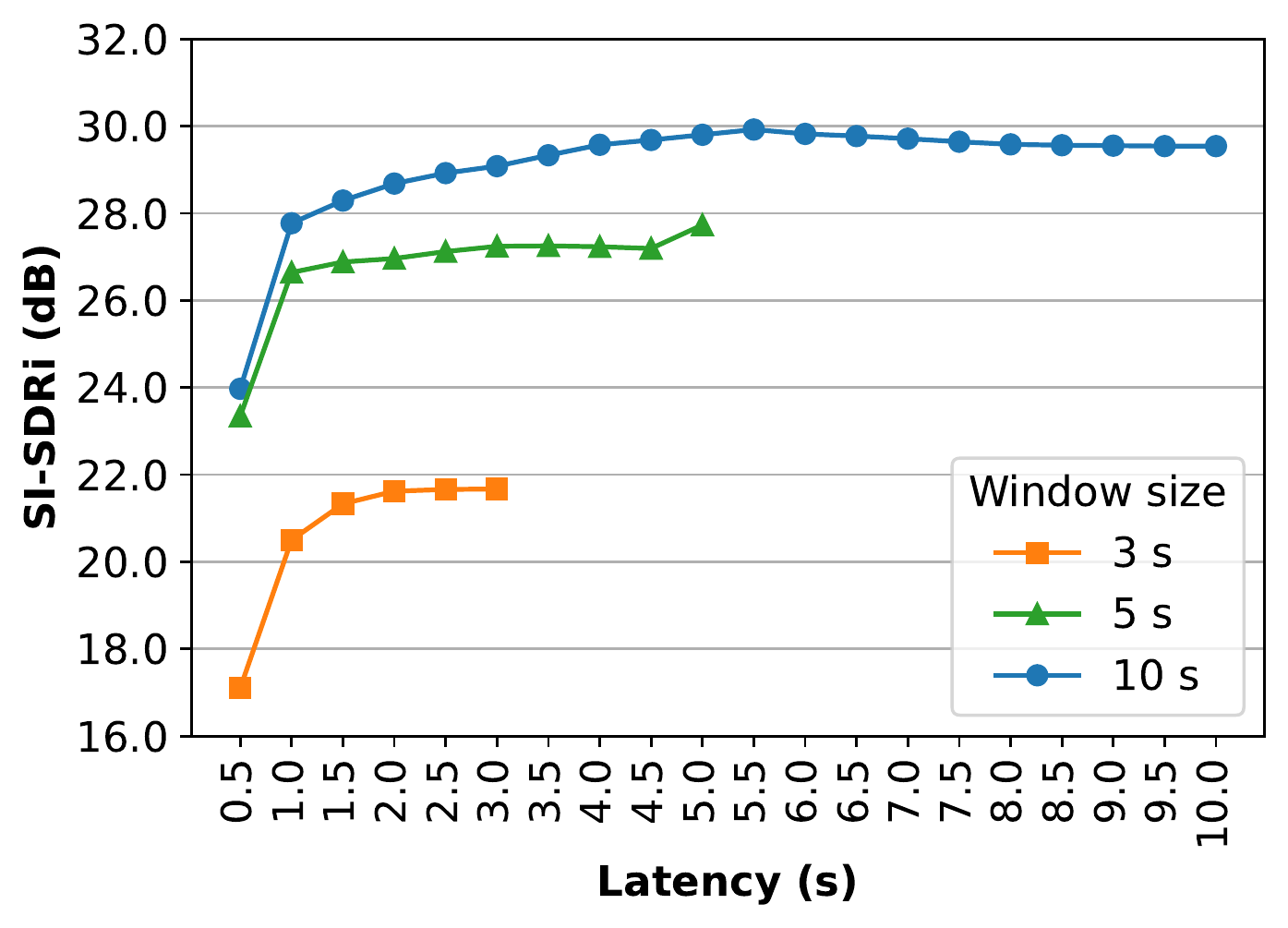}
  \caption{Overlap ratio: $10$\%}
  \label{fig:res/causal_analysis/10}
\end{subfigure}
\begin{subfigure}[b]{.5\textwidth}
  \centering
  \includegraphics[width=1.\textwidth]{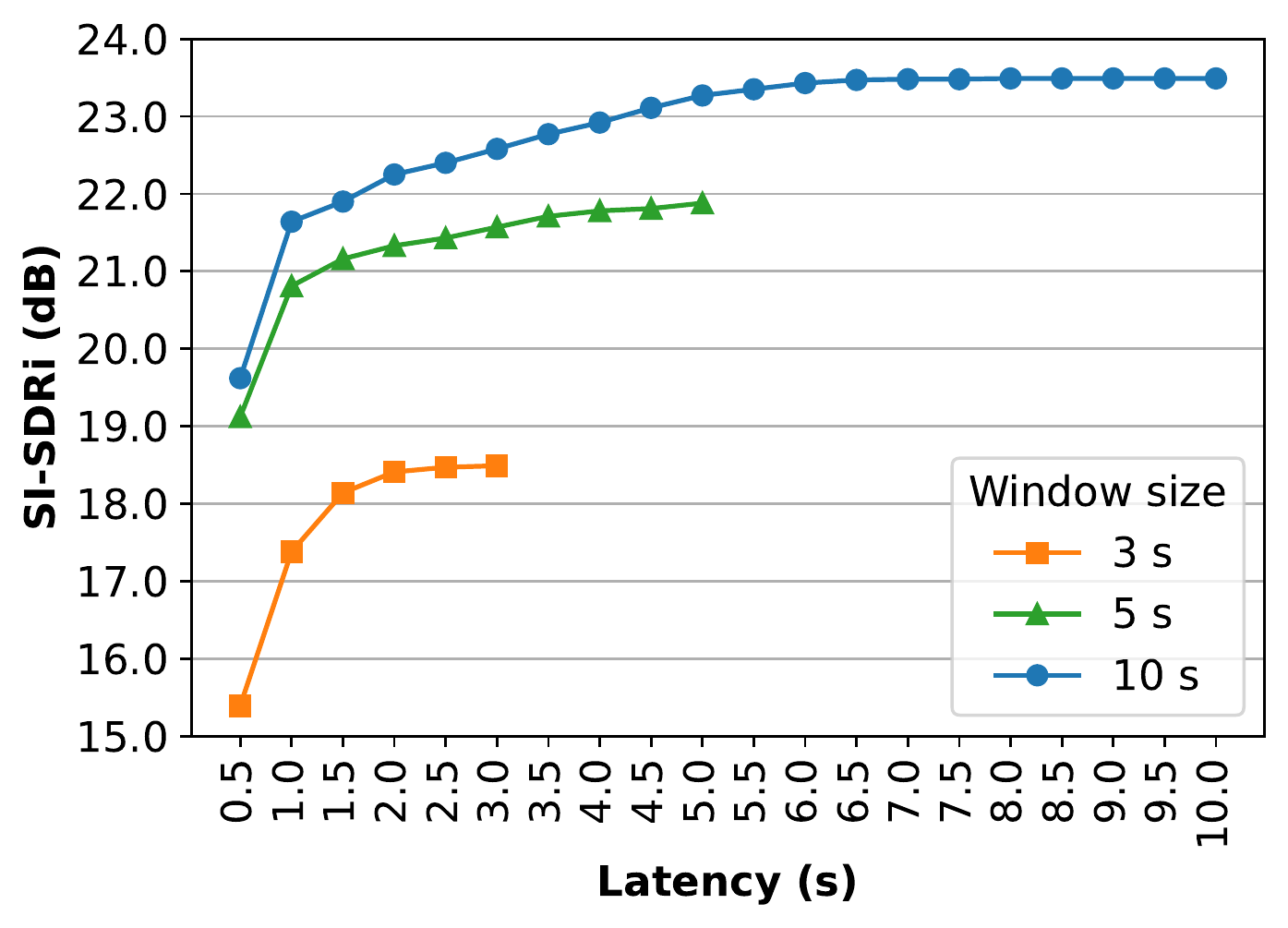}
  \caption{Overlap ratio: $20$\%}
  \label{fig:res/causal_analysis/20}
\end{subfigure}
\begin{subfigure}[b]{.5\textwidth}
  \centering
  \includegraphics[width=1.\textwidth]{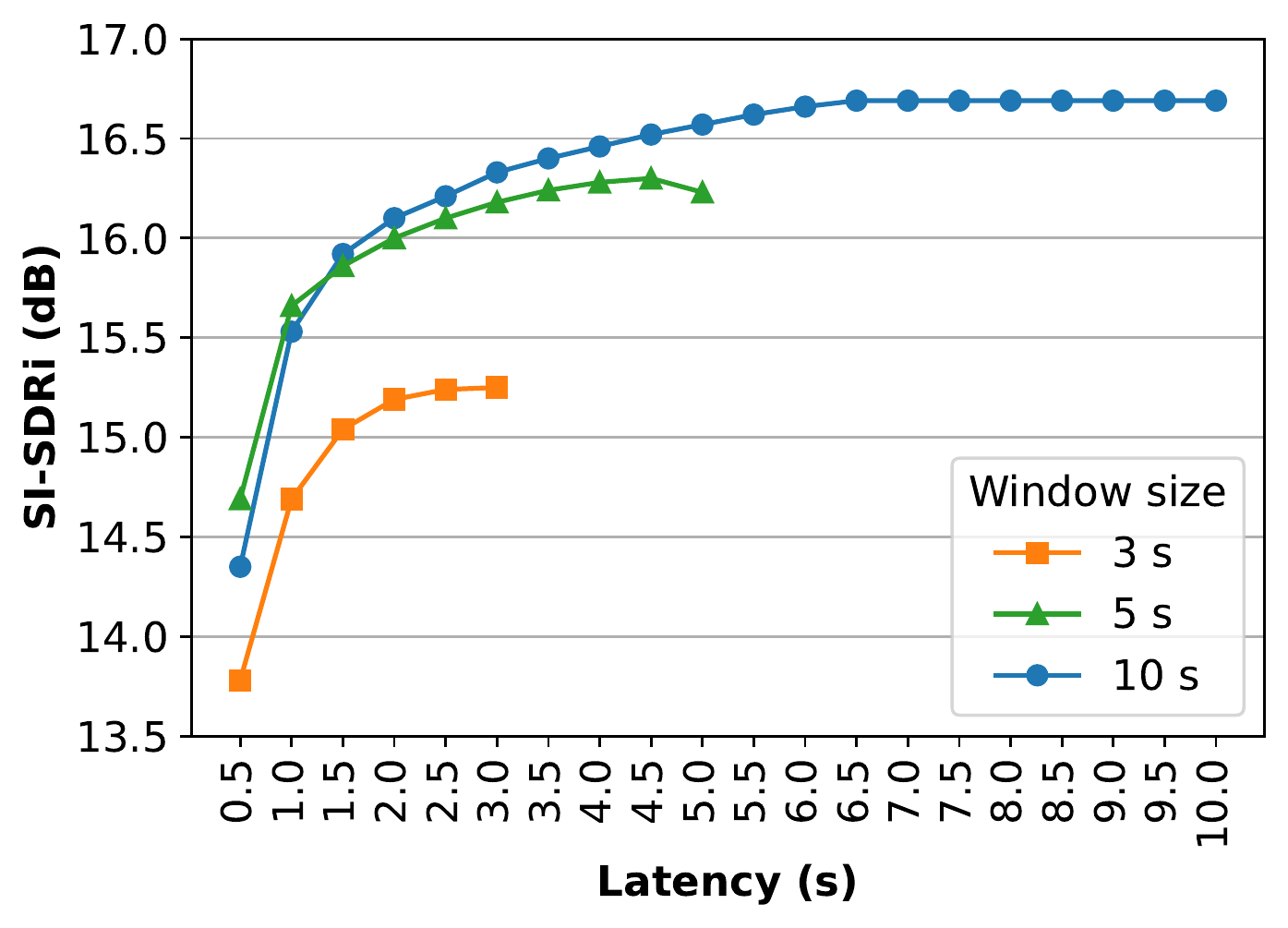}
  \caption{Overlap ratio: $40$\%}
  \label{fig:res/causal_analysis/40}
\end{subfigure}
\begin{subfigure}[b]{.5\textwidth}
  \centering
  \includegraphics[width=1.\textwidth]{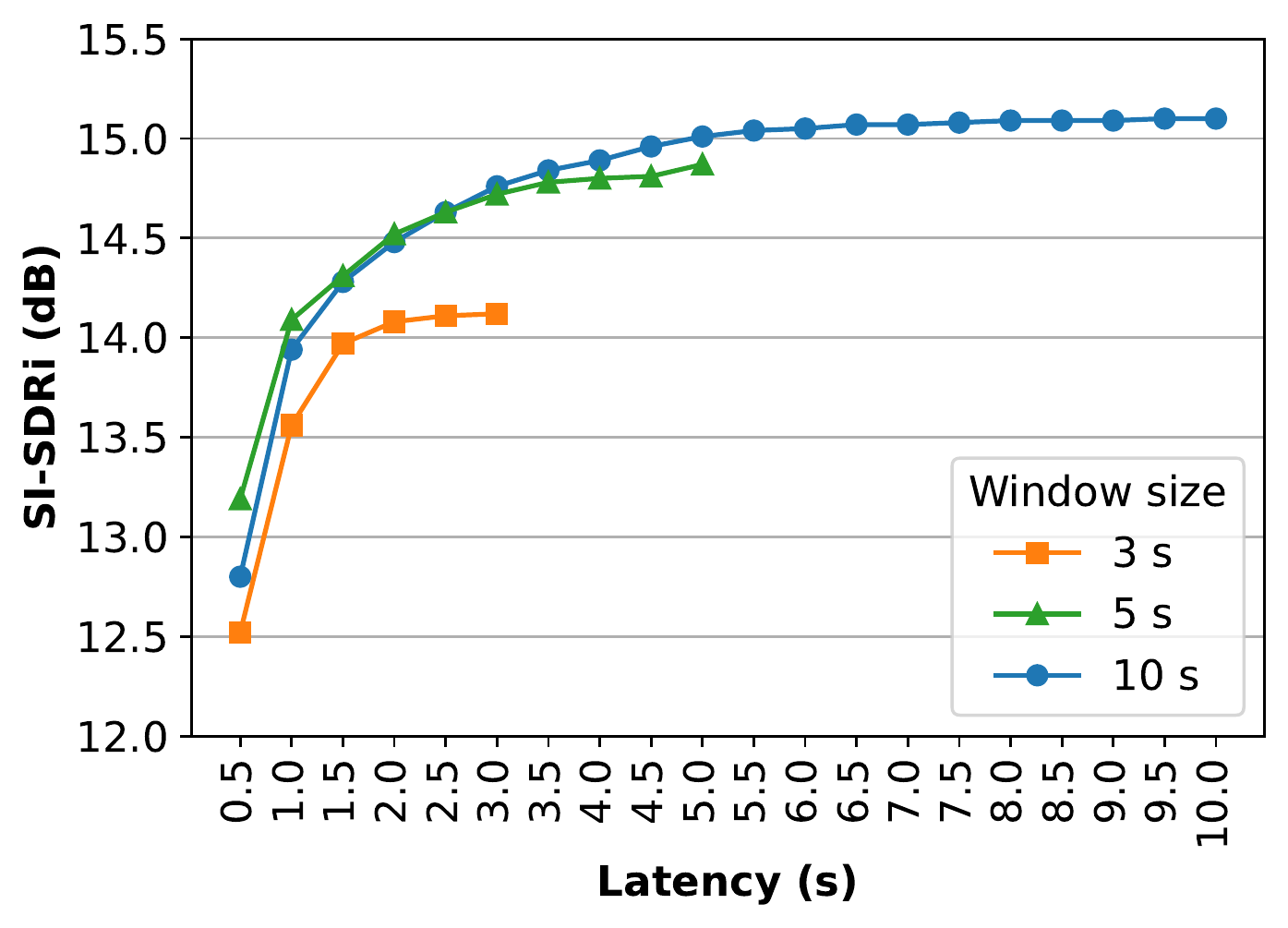}
  \caption{Overlap ratio: $60$\%}
  \label{fig:res/causal_analysis/60}
\end{subfigure}
\hfill
\begin{subfigure}[b]{.5\textwidth}
  \centering
  \includegraphics[width=1.\textwidth]{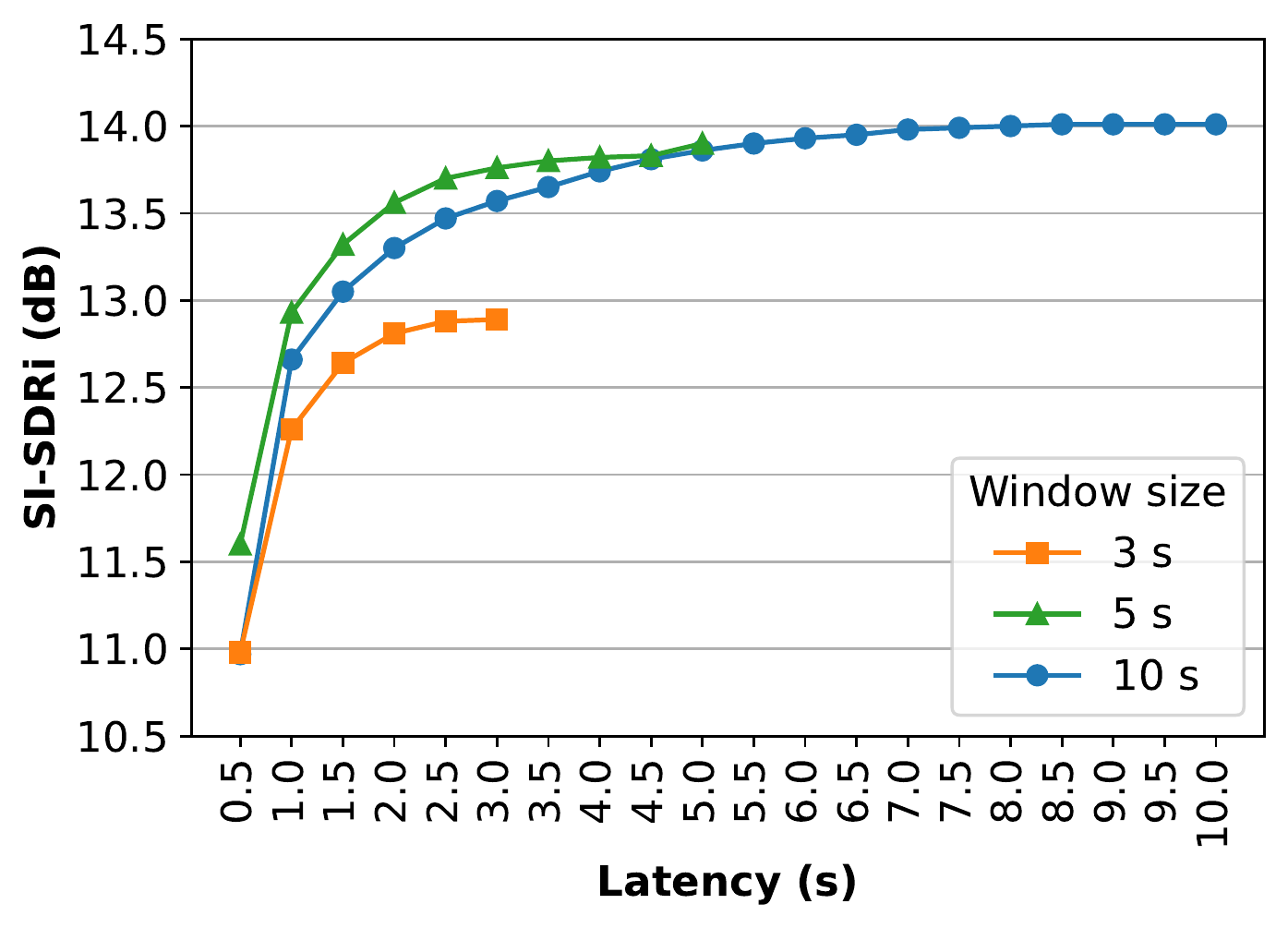}
  \caption{Overlap ratio: $100$\%}
  \label{fig:res/causal_analysis/100}
\end{subfigure}
\vspace*{-2.5mm}
\caption{Results on sparse test sets for different latencies. The hop size for consecutive frames is set to $0.5$ s. The number of segments used to generate the output is equal to the ratio between the latency and the hop size.}
\label{fig:res/causal_analysis}
\end{figure*}

\subsection{Online/Causal Evaluation}
\label{ssec:res/online_eval}

We perform an analysis of using an offline speech separator in a causal setting, as described in the Subsection \ref{ssec:sys_descr/online_proc}. We evaluate the fine-tuned CSS system (i.e., Model $2$ in Table \ref{tab:res/sep_win}) by only exploiting some of the output estimations for a given segment of length $H$.

In Fig. \ref{fig:res/causal_analysis} are shown the results on the sparse test set for different processing window lengths (i.e., $3$, $5$ and $10$ s) and for several overlap ratios. The length of each curve is different since the number of possible estimations and latencies depends on the window size, and can range from 1 to $t_W$. We set the hop length to $0.5$ s which is short enough to enable online processing for many applications (e.g. real-time captioning) and suited to carry out an interesting comparison between the different windows when varying latency. Since we are using bidirectional recurrent models as building blocks in the DPRNN, we need to process the entire frame at each hop, resulting in higher resource usage for larger windows. The computational cost is of $76$ G, $127$ G and $254$ G FLOPS for $3$-, $5$- and $10$-seconds long windows, respectively. As the DPRNN consists of convolutional and BLSTM layers, then the maximum GPU memory usage theoretically grows linearly with the window size. In practice, the memory usage depends on the specific algorithmic implementation. This usually leads to higher peak memory allocation than theoretical one. In our case, we use the PyTorch \cite{paszke2019pytorch} framework to estimate the GPU memory footprints which are of $0.94$, $2.07$ and $4.06$ GB for windows of $3$, $5$ and $10$ s, respectively. 

As expected, SI-SDR generally improves when latency and window size increase. For offline processing, as there are no latency and resource usage constraints are not very strict, the largest window with maximum latency can be used. However, in most real-world applications both latency and resource usage are very important and can not be neglected.

Another clear finding is that using a single segment degrades a lot the performance for all windows and speech overlap conditions. This case corresponds to the lowest latency setting. We output segments of length $H$ at the right edge of the frames - see $Y_1$ in Fig. \ref{fig:sys_descr/win_seg} - which are less accurate since they do not take advantage of information from future segments. Therefore, this configuration should be selected only when low-latency is very important and perfect separation is not necessary.

For the same latency, the $3$-seconds long window performs much worse compared to the longer windows. However, as its separation capability is still decent even in the worst case (i.e., $10.98$ dB for $100$\% overlap and $0.5$ s latency), it could be a convenient choice for low-power consumption hardware (e.g., mobile phones and wearable devices). On the other hand, the separation capability of the systems based on the $5$- and $10$-seconds long windows is very similar for high overlap ratios (i.e., $>40$\%). In some cases, the $5$-seconds window provides better results. In particular, in the $100$\% overlap ratio scenario it always outperforms the $10$-seconds long window. If we are only interested in separating high-overlapped speech with relatively low latency, i.e., below $5$ s, choosing a $5$-seconds long window is a good solution since we can obtain the same performance by halving computational cost and memory footprint. However, the largest window (i.e., $10$ s) outperforms by large margin for low-overlap ratios, which are the most common scenarios in meetings and natural conversations. We observe that when the SI-SDR improvement is above $20$ dB the separation in most cases is almost perfect and outputs indistinguishable from original sources. The values for the $5$-seconds long window are all above this threshold, then it is generally a good trade-off for each overlap ratio. 

\section{Conclusions}
\label{sec:conclusions}

Streaming speech applications require causal and low-latency processing, while keeping power and resource consumption low.
In this paper, we investigated the use of an offline separation algorithm, i.e., DPRNN \cite{luo2020dual}, for continuous speech separation in natural conversations. We tested both a model trained on fully overlapped speech only and another model fine-tuned on sparsely overlapped speech. Results showed that longer processing windows provided better performances for each speech overlap ratio. The fine-tuned system improved a lot the SI-SDR metric, especially for low overlapped test sets, which represent the most natural scenario in real-world environments (e.g., meeting and phone conversations). Additionally, we observed that using a very short window, i.e., $1$ s, in the framing phase failed at separating long mixtures. The main cause relied on low accuracy in reordering consecutive frames using a similarity metric based on cross-correlation.

Then, we presented a framework that enabled to convert an offline separation algorithm into an online/causal one, which can be used to process input audio streams with a desired minimum latency. We performed an extensive comparison between several processing windows in different speech overlap conditions. We found that using a $5$-seconds long window was in general a good trade-off between separation performance and resource consumption.

We believe that this work can provide the speech processing community with some insights on how to choose the best CSS configuration (i.e., window length and latency) for different applications. 

Future works might compare the CSS framework with a speech separation model natively designed to process audio streams in an online fashion. These models, perform worse but, since they employ causal operations only (e.g., no bidirectional recurrent layers), they do not require to process the entire CSS window at each hop. That results in great computational and memory savings while retaining low-latency. A comparison of native and non-native causal CSS systems could provide a more comprehensive overview over the possible design choices and trade-offs for an effective CSS system.
Another possible future direction could be testing other similarity functions used to reorder consecutive frames. In this study we employed the cross-correlation function, which performed very poorly with short frames, i.e., $1$ s. That imposed a lower bound on the minimum resource requirements needed to guarantee an adequate level of accuracy. The results obtained with the oracle information (cf. Table \ref{tab:res/sep_win_oracle}) suggested that there was a large room for improvement with short frames. Alternative similarity functions could be based on speaker embeddings (e.g., \cite{snyder2018x, desplanques2020ecapa}) or even learnt from training data.

\section{Acknowledgements}
This work has been supported by the AGEVOLA project (SIME code 2019.0227), funded by the Fondazione CARITRO.

\bibliography{refs}
\bibliographystyle{IEEEtran}

\end{document}

%% file: math_commands.tex
\usepackage{amsmath,amsfonts,bm}

\def\eqref#1{equation~\ref{#1}}

\def\1{\bm{1}}

\DeclareMathAlphabet{\mathsfit}{\encodingdefault}{\sfdefault}{m}{sl}
\SetMathAlphabet{\mathsfit}{bold}{\encodingdefault}{\sfdefault}{bx}{n}

